\documentclass[12pt]{article}

\usepackage[latin1]{inputenc}
\usepackage[T1]{fontenc}
\usepackage[english]{babel}
\usepackage{epsfig}
\usepackage{amssymb}
\usepackage[centertags]{amsmath}
\usepackage{amsthm}
\usepackage{amsfonts}
\usepackage{makeidx}
\usepackage{graphics}
\usepackage{graphicx}
\usepackage{psfrag}
\usepackage{subfigure}
\usepackage{dsfont}
\usepackage{eucal}
\usepackage{color}
\usepackage{bm}
\usepackage[square,comma,sort&compress]{natbib}
\usepackage{fancybox}
\usepackage{footnpag}
\usepackage{latexsym}
\usepackage{fancyhdr}
\usepackage{textcomp}
\usepackage{latexsym}
\usepackage{fancyhdr}
\usepackage[marginal]{footmisc}
\usepackage{dsfont}

\makeindex

\def \de {\partial}
\def \l {\lambda}

\def \D {\Delta}
\def \b {\beta}
\def \a {\alpha}

\def \h {\eta}

\def \D {\Delta}
\def \e {\varepsilon}

\def \r {\rho}

\def \m {\mu}
\def \n {\nu}

\def \non {\nonumber}
\def \noi {\noindent}
\def \ra {\rightarrow}
\def \xra {\xrightarrow}

\def \pr {\prime}

\def \miu {\leqslant}

\def \fr {\displaystyle\frac}

\def\laq{~\raise 0.4ex\hbox{$<$}\kern -0.8em\lower 0.62
ex\hbox{$\sim$}~}
\def\gaq{~\raise 0.4ex\hbox{$>$}\kern -0.7em\lower 0.62
ex\hbox{$\sim$}~}

\def \wt {\widetilde}

\begin{document}
\thispagestyle{empty} \vspace*{1cm} \rightline{BARI-TH/07-562}
\vspace*{2cm}
\begin{center}
  \begin{LARGE}
  \begin{bf}
On the light glueball spectrum \\ 
\vspace*{0.2cm}
in a holographic description of  QCD
\vspace*{0.5cm}
  \end{bf}
  \end{LARGE}
\end{center}
\vspace*{8mm}
\begin{center}
\begin{large}
P.~Colangelo$^a$, F.~De~Fazio$^a$, F.~Jugeau$^{a}$ and S.~Nicotri$^{a,b}$
  \end{large}
\end{center}
\vspace*{8mm}
\begin{center}
\begin{large}
\begin{it}
$^a$Istituto Nazionale di Fisica Nucleare, Sezione di Bari,
Italy\\ $^b$ Universit\`a degli Studi di Bari, Italy
\end{it}
  \end{large}
\end{center}
\begin{quotation}
\vspace*{1.5cm}
\begin{center}
  \begin{bf}
  Abstract\\
  \end{bf}
  \end{center}
\noi
We investigate the spectra  of light  scalar and vector glueballs in
a holografic description of QCD   with a dilaton background bulk
field. In particular, we study how the glueball masses depend on the conditions on the dilaton background  and on the geometry of the bulk. 

\end{quotation}

\newpage
\setcounter{page}{1}

\section{Introduction}
A breakthrough in the attempt to understand strongly coupled
Yang-Mills theories is represented by the AdS/CFT correspondence conjecture, stating 
that a connection can be established between the
supergravity limit of a superstring/M-theory living on a $d+1$ anti
de Sitter (AdS) space times a compact manifold  and the large $N$ limit of
a maximally $\CMcal{N}=4$ superconformal
 $SU(N)$ gauge theory defined in the $d$ dimension AdS boundary
\cite{Maldacena:1997re,Witten:1998qj,Gubser:1998bc,Aharony:1999ti}. However, the  application of this conjecture 
to a theory such as QCD is not straightforward, being  QCD neither
supersymmetric nor  conformal. Witten  proposed a procedure to extend  duality
to such gauge theories \cite{Witten2:1998}: the conformal invariance
is broken by compactification (the compactification radius giving
rise to a dimensionful   parameter, namely the mass gap of QCD),
while supersymmetry is broken by appropriate boundary conditions on
the compactified dimensions. The AdS geometry of the dual  theory is
then deformed into an AdS-black-hole geometry where the horizon
plays the role of an IR brane.   In this  (so-called up-to-bottom)   approach, analyses of the
 glueball spectrum have been  carried out, obtaining, for example, that the operator $Tr F^2$ in four
 dimensions corresponds to the massless  dilaton field in supergravity in ten dimensions,   that the scalar glueball  with $J^{PC}=0^{++}$ in QCD is related to the dilaton propagating in the black-hole geometry and its mass is computable by solving the dilaton wave equation   \cite{Ooguri, Evans2,Evans:2005ip}.  The  numerical results are   close to the available lattice data \cite{lattice}.
 
However,  one could
adopt the strategy of  investigating which features the dual
theory should have in order to reproduce known QCD properties. In
this (so-called bottom-up) approach, instead of trying to deform the high dimensional theory
to obtain a theory in $4d$ with similarities with QCD,
one begins with QCD and attempts to construct a five dimensional holographic dual. A hint to follow
is that, although QCD is not itself a conformal theory, it nevertheless resembles  a strongly coupled conformal theory  in the domain  where the quark masses are neglected and the coupling is approximately constant   (the possibility that the  QCD $\beta$ function has
an infrared fixed point is discussed, e.g., in \cite{QCDtheory,QCDpheno}). As  pioneered by  Polchinski and Strassler,  it is possible to implement duality in these nearly conformal conditions 
definining QCD on the four-dimensional boundary, and  introducing a bulk space which is a slice $AdS_5$, the  size of which stands for an IR cutoff associated to the QCD mass gap, the
so-called hard IR wall approximation  \cite{polchinski}. 
This procedure was investigated in
refs.\cite{deTeramond,son,altri0,altri,braga} with the calculation of the light hadron spectrum. Moreover,
the glueball spectrum was studied considering   various boundary conditions of the associated $5d$ field  at the IR brane \cite{braga}.  
The static $Q \bar Q$ potential was also  worked out  \cite{Andreev:2006eh} together with
hadron   wavefunctions and form factors \cite{wavef}.
Besides,  leaving the hard IR wall and considering a background dilaton field, it was shown that  
 properties of QCD can be reproduced, namely  the Regge behaviour of light mesons \cite{son2}, 
 at odds with what happens starting from a general string
theory and attempting to deform it \cite{shifman}. 

Even though in
the bottom-up approach
 $\alpha_s$ corrections,  running of the coupling constant,
 geometry of the compact manifold which should be considered together with $AdS_5$,   etc., at present are left aside, there is the  hope that   the main features of the dual theory can be identified.

 The starting point is inspired by a principle of the AdS/CFT correspondence,
which establishes 
a one-to-one correspondence between a certain class of local
operators  (namely, the chiral primary operators and
their superconformal descendants) in the $4d$ $\CMcal{N}=4$
superconformal gauge theory and supergravity fields 
representing the holographic correspondents in the $AdS_{5}\times
S^{5}$ bulk theory \cite{Witten:1998qj,Gubser:1998bc,  Aharony:1999ti}.   Analogously, in the bottom-up approach  one attempts to construct a correspondence 
between QCD local operators and fields in the $AdS_5$ bulk space.  Although
in this way  the five dimensional dual of QCD contains
an infinite number of fields, in correspondence to the infinite
number of QCD operators,    it was shown
that,  considering only few operators relevant for chiral dynamics, a few  properties in the light meson sector can be obtained,  namely the $\r$ meson spectrum, 
the axial-vector meson spectrum, the $\r \pi \pi$ coupling and   a few  leptonic 
constants,  with a small  number of hadronic parameters  .

 In this note  we consider the spectrum of low-lying scalar and vector glueballs   in the  approach,  proposed in \cite{son2},  where the hard
 IR cutoff in the $AdS_5$ space  is replaced by a smooth cutoff  obtained  introducing a 
 dilaton  background bulk  field.   In particular, we study how the glueball masses depend on the
 conditions on the background dilaton  and on  the bulk geometry, so that they can be compared to, e.g.,  
lattice QCD or QCD sum rule results \cite{lattice,narison}.

\section{Model for a $5d$ holographic dual of QCD}
Following \cite{son2},  we consider  a five dimensional conformally flat spacetime (the bulk) described by the metric
\begin{eqnarray}
g_{MN}=e^{2A(z)}\h_{\,MN} &,& ds^2=e^{2A(z)}(\h_{\,\m\n}dx^\m
dx^\n+dz^2)
\end{eqnarray}
($M,N=0,\dots 4$), where $\h_{\,MN}=\hbox{diag}(-1,1,1,1,1)$,
$x^\mu$ ($\mu=0,\dots 3$)  represent the usual space-time  (the
boundary) coordinates and $z$ is the fifth holographic coordinate
running from zero to infinity. The metric function $A(z)$ satisfies
the condition
\begin{equation}\label{UV}
A(z)\xra[z\ra0]{}\ln \biggl(\frac{R}{z}\biggr)
\end{equation}
 to reproduce the $AdS_5$ metric close  to the UV brane $z \to 0$;
in the following  we put  to unity the   radius $R$. Besides,  we consider a background dilaton
 field $\phi$ which only depends on the holographic coordinate $z$ and vanishes at the UV brane.  By an appropriate choice of  the $\phi$ dependence in the
 IR (large values of $z$) we  construct a $5d$ model that can   be considered similar to a cutoff AdS space:
a smooth cutoff in the IR replaces the hard-wall IR cutoff that would be obtained by allowing
 the holographic variable $z$ to vary  to a maximum value 
 $\displaystyle z_m\simeq\frac{1}{\Lambda_{QCD}}$. The introduction of a background dilaton
 allows to avoid ambiguities in the choice of the field boundary conditions at the IR wall.

To investigate the mass spectra of the QCD  scalar  
and vector  glueballs, we  consider the
two lowest dimension operators with the corresponding quantum
numbers and defined in the field theory living on the $4d$ boundary:

\begin{equation}\label{boundoper}
  \begin{cases}
O_S= Tr(F^2)\\
O_V =Tr(F(DF)F)
  \end{cases}
\end{equation}
(with $D$ the covariant derivative) 
 having conformal dimension $\D=4$ and $\D=7$, respectively.  The operator corresponding to the vector glueball satisfies the Landau-Pomeranchuk-Yang selection rule \cite{shifman1}. 
  In the  AdS/CFT correspondence  the conformal dimension
 of a ($p-$form) operator on the boundary  is related to the $(AdS\;mass)^{2}$ of its  dual field in the
 bulk as follows \cite{Witten:1998qj,Gubser:1998bc}:

\begin{equation}\label{m5}
(AdS\;mass)^2=(\D-p)(\D+p-4)\;\;.
\end{equation}
In the following we assume that the mass $m_5^2$ of the bulk fields is given by this expression. 
 
A $5d$ massless  scalar field $X(x,z)$ can be constructed as the  correspondent of  $Tr F^2$, described by  the  action in the gravitational background:

\begin{equation}\label{actionscalmass}
S=-\fr{1}{2}\int d^5x\,\sqrt{-g}\,e^{-\phi(z)}\,\,  g^{\,MN}(\de_{M}X) (\de_{N}X) \,\,\,\, 
\end{equation}
with  $g=det(g_{MN})$.
Scalar glueballs  are identified as the normalizable modes of   $X$
 satisfying the equations of motion obtained from \eqref{actionscalmass},  corresponding
to a finite action. 

For the spin 1 glueball,   we introduce a 1-form $A_M$ described by the action:
\begin{equation}\label{actionvecmass}
S=-\fr{1}{2}\int d^5x\,\sqrt{-g}\,e^{-\phi(z)}\,\biggl[ \fr{1}{2}
g^{\,MN}g^{\,ST} F_{MS} F_{NT}+m_5^2\,g^{\,ST}A_SA_T\biggr]
\end{equation}
with $F_{MS}=\de_{M} A_S - \de_{S} A_M$ and  $m_5^2=24$,
and study  its normalizable modes. 
Notice that the action \eqref{actionvecmass}, with a different value of $m_5^2$,  describes
fields that are dual to other operators in QCD,  namely those describing 
 hybrid mesons with spin one,  which is  an explicit example of different QCD operators having  similar bulk fields as holographic correspondents.  
 
 In the following Section we discuss how the spectrum  can be worked out. 
 However,  before such a discussion,  it is interesting to  comment on the pseudoscalar glueball, described in QCD by the $\D=4$ operator
$O_P=Tr (F \wedge F)$. Identifying another  0-form in the bulk as the  correspondent of $O_P$ and
 describing it  by the same action \eqref{actionscalmass}, 
a degenerate mass spectrum would be obtained  for  scalar and 
pseudoscalar glueballs, at odds with the results obtained, e.g., in lattice QCD where it is found that
the mass of the lightest scalar glueball is smaller than the mass of the lightest pseudoscalar one
\footnote{A parity degeneracy has 
been pointed out  in the light
baryon  spectrum  in the framework of a holographic dual of QCD  \cite{deTeramond}.}.  A way out to such a degeneracy issue could be represented by the choice of  still considering
 the relation \eqref{m5} (keeping in mind that such a relation rigorously  holds only in the AdS/CFT correspondence conditions), and attempting 
 to  describe also the field  corresponding to  the pseudoscalar glueball by a massive 1-form $A_M$.
 There is an indication  for that,  since in the top-to-bottom approach
 the pseudoscalar glueball is described by a massless R-R 1-form
\cite{Evans:2005ip}.  The construction followed in that approach is that,  after a first compactification of an $11d$ M-theory in $AdS_{7}\times S^{4}$, AdS/CFT establishes a
duality between a Type IIA string theory in $AdS_{6}\times S^{4}$
and a low-energy effective supersymmetric theory $SU(N)$ described
in terms of N coincident D4-branes. On this $5d$ D4-brane
worldvolume, the field that couples to $TrF\wedge F$ is a massless
 R-R 1-form  $A_\r$,  the coupling term reading as  \cite{Aharony:1999ti}
\begin{equation}
  \fr{1}{16\pi^2}\int d^5x\,\e^{\r\m\n\a\b}A_\r
  F_{\m\n}F_{\a\b}\;\;.
\end{equation}
Then, a second compactification provides a non-supersymmetric model
of QCD in terms of N coincident D3-branes dual to the Type IIA
string theory in an AdS-Black-Hole geometry. The mass spectrum of
the pseudoscalar glueball is then determined by solving the equation
of motion for the  R-R 1-form in the $10d$
bulk\footnote{The spectrum of
the pseudoscalar glueball has also  been analyzed  using a  massive 3-form of the $11d$
supergravity coupled to a $\Delta=9$ operator of the $6d$ boundary
theory \cite{Ooguri}.}.  We do not continue here in such an analysis, since the issue of parity of various hadronic excitations deserves a dedicated study.

\section{Background fields}
The  metric function $A(z)$  and the background dilaton field $\phi(z)$  can be constrained.  A constraint to $A(z)$ is 
 the condition \eqref{UV} which, together with
 the requirement   $\displaystyle \phi(z) \xra[z\ra0]{}0$,  allows to reproduce the $AdS_5$ metric close to UV brane $z\simeq 0$. On the other hand, 
  a suitable  large 
 $z$ dependence of  $\phi(z)$ can be fixed 
 to reproduce the Regge behaviour of the low-lying mesons.
The two conditions

\begin{equation}\label{constraints}
  \begin{cases}
    \phi(z)-A(z)\xra[z\ra0]{}\ln z\\
    \\
    \phi(z)-A(z)\xra[z\ra\infty]{}\,z^2\\
  \end{cases}
\end{equation}
satisfy the two  requirements: indeed,   the first condition  
satisfies eq.\eqref{UV},   while the second  one  allows  to
recover the Regge behaviour of $\r$   resonances, as shown in  \cite{son2}. Moreover,  the
metric  function $A(z)$ must not have any contribution growing as
$z^2$ at large $z$, a condition coming from computing   the masses  of higher spin
mesons  \cite{son2}.

The simplest choice consistent with these constraints\footnote{We put to one  the scale parameter multiplying $z^2$ in  the dilaton field;   mass predictions  will be given in units of this parameter.}
\begin{eqnarray}\label{choice}
\phi(z)&=&z^2 \non \\
A(z)&=&-\ln z
\end{eqnarray}
has been chosen to calculate  the  spectrum  of mesons of spin  $S$ and
radial quantum number $n$, with the result: $m^2_n=4(n+S)$  \cite{son2}. 

We use these expressions  for the background dilaton and the  metric function to
work out  the glueball spectrum.   The
field equations of motion obtained from the actions
(\ref{actionscalmass})-(\ref{actionvecmass}) can be reduced in the
form of a one dimensional Schr{\"o}dinger equation in the variable
$z$:

\begin{equation}\label{geneqpot}
  -\psi^{\pr\pr}+V(z)\psi=-q^2\psi
\end{equation}
involving the function $\psi(z)$ obtained
 applying a  Bogoliubov  transformation
$\psi(z)=e^{-B(z)/2}\wt{Q}(q,z)$  to the Fourier transform
$\wt{Q}$  of the field $Q$ ($Q=X, A_M$) with respect to the boundary variables 
 $x^\m$. The function $B(z)$ is a combination of the dilaton
and the metric function: $B(z)=\phi(z)-c \, A(z)$,  with the
parameter $c$ given by:  $c=3$  and  $c=1$
 in  cases of $X$ and $A_M$ fields,  respectively. 
The condition $q^2=-m^2$ identifies the mass of  the normalizable modes of the two fields. 
 
Eq. \eqref{geneqpot}  is a one dimensional Schr{\"o}dinger equation where $V(z)$ plays  the role of a potential. It  reads as:

\begin{equation}\label{potscalar}
 V(z)=\fr{1}{4}(B^\pr(z))^2-\fr{1}{2}B^{\pr\pr}(z)+\fr{m_5^2}{z^2}=V_0(z)+\fr{m_5^2}{z^2}
\end{equation}
with
\begin{equation}\label{potentials}
V_0(z)= z^2+\frac{c^2+2c}{4 z^2}+c-1 \,\,\, .
\end{equation}
With this potential  eq.\eqref{geneqpot} can be  analytically solved. Regular solutions at
$z\to 0$ and $z \to \infty$ correspond to  the spectrum:

\begin{equation}\label{specgenscal}
  m_n^2=4n+1+c+\sqrt{(c+1)^2+4m_5^2}
\end{equation}
with $n$ an integer (we identify it as a radial quantum number), while   the 
corresponding eigenfunctions read as:

\begin{equation}\label{kummerscal}
  \psi_n(z)=A_n\,e^{-z^2/2}z^{g(c,m_5^2)+1/2}\;
  {_1F_1}\left(-n,g(c,m_5^2)+1,z^2\right)
\end{equation}
with  ${_1F_1}$ the Kummer confluent hypergeometric
function,  $A_n$
a normalization factor,  and $g(c,m_5^2)=\sqrt{\frac{(c+1)^2}{4}+m_5^2}$.
From these relations  we obtain the spectrum of  scalar and vector  glueballs:
\begin{eqnarray}
m_n^2&=&4n+8\;\;  \label{specscal}\\
m_n^2&=&4n+12\;\;  \label{specvec}
\end{eqnarray}
respectively.

A few remarks are in order.  First, both the spectra  have the same dependence on the radial quantum
number $n$ as the  mesons of spin $S$:  this is a consequence of the large $z$ behaviour chosen for
the background dilaton.
 Second, both the lowest lying gueballs are heavier than
the $\r$ mesons, the spectrum of which reads:  $m_n^2=4n+4$,  as derived in \cite{son2}.
 Finally,  the vector  glueball turns out to be  heavier than the scalar one.
  
 Comparing our result to the computed $\r$ mass, we obtain for the lightest scalar   $(G_0)$  and vector $(G_1)$    glueballs
\begin{equation}\label{latticeres}
\frac{m^2_{G_{0}}}{m^2_\r}=2 \hspace*{2cm}
\frac{m^2_{G_{1}}}{m^2_\r}=3
\end{equation}
which implies that these  glueballs are expected to be lighter than as predicted by other QCD approaches \cite{lattice}. Moreover, the result $m^2_{G_1}-m^2_{G_0}=m^2_\r$ predicts a lightest
vector glueball with mass below 2 GeV.
 
It is interesting to investigate how it is possible to modify   the $z$
dependence of the background dilaton field and of the metric function $A$, and how the spectra change, an issue  discussed in the following Section.

\section{Perturbed background}
There are other  choices for the background dilaton $\phi$ and the metric function $A$ which
satisfy the constraints in \eqref{constraints} and may  modify the predictions for the
 scalar and vector glueball masses.
As a matter of fact, it is 
possible to add to the background fields terms of the type $z^\a$
with $0\miu\a<2$. Considering  the simplest case:
$\a=1$,  this can be done in   two different
ways.  Instead of using \eqref{choice},  we can modify the dilaton
field including a linear contribution which is subleading in the IR regime
$z \to \infty$:
\begin{eqnarray}\label{moddil}
\phi(z)&=&z^2+\l z \non\\
A(z)&=&-\ln z
\end{eqnarray}
 with $\l$ a real parameter.
 Another  possibility consists in modifying  the metric function,
\begin{eqnarray}\label{modgeo}
\phi(z)&=&z^2 \non \\
A(z)&=&-\ln z- \l \,  z
\end{eqnarray}
which now acquires a linear  term subleading in the  UV regime $z
\to 0$. The two choices  produce different results.

\begin{figure}[h]
 \begin{center}
\includegraphics[width=0.45\textwidth] {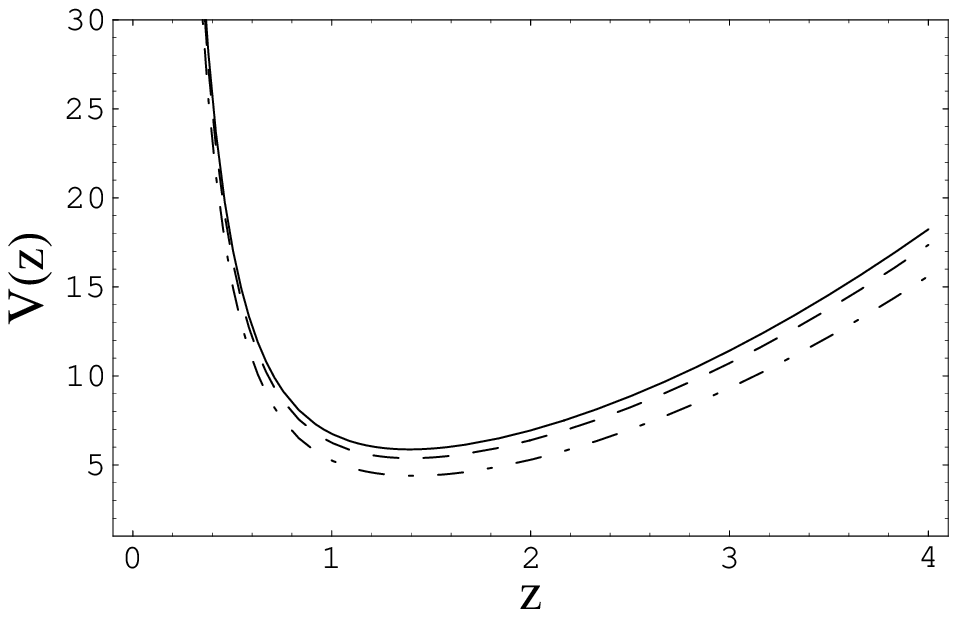}
\includegraphics[width=0.45\textwidth] {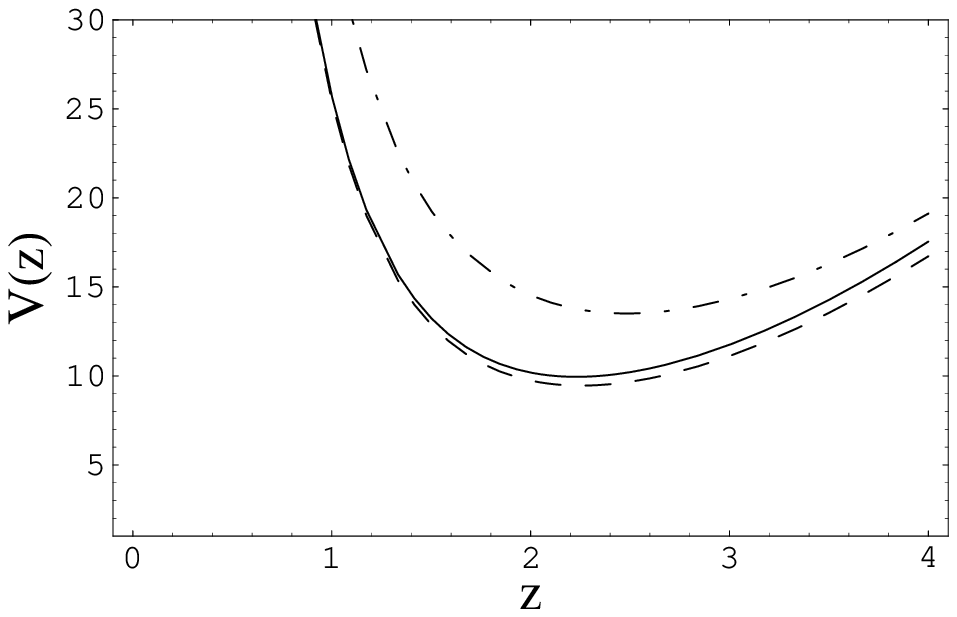}
 \end{center}
 \caption{\footnotesize{The unperturbed potential (solid line), and the potential obtained perturbing
 the dilaton field  (dashed line) and the metric (dot-dashed line)
 for scalar (left) and vector glueball (right) using $\l=-0.2$.}}\label{fig:scalfig}
\end{figure}

Using  the expressions \eqref{moddil},  i.e. modifying the dilaton
field,   the potential \eqref{geneqpot} becomes:
\begin{equation}\label{pert-pot}
V(z)=V_0(z) +\lambda V_1(z)+\frac{\lambda^2}{4}+\frac{m_5^2}{z^2}
f(z,\lambda)
\end{equation}
with $V_0(z)$  given in eq.\eqref{potscalar} and
\begin{eqnarray}\label{v1}
V_1(z)&=&z+\frac{c}{2z} \non \\
f(z,\lambda)&=&1 \,\,\,\, .
\end{eqnarray}
On the other hand, using the expressions in \eqref{modgeo},  i.e.
modifying the metric  in the IR,  the potential term reads as:
\begin{equation}\label{pert-pot-1}
V(z)=V_0(z) +\lambda \tilde V_1(z)+{c^2 \lambda^2 \over 4}+{m_5^2 \over z^2}
f(z,\lambda) \,,
\end{equation}
where
\begin{eqnarray}\label{v1-1}
\tilde V_1(z)&=&c \left( z+{c \over 2z} \right) \non \\
f(z,\lambda)&=&e^{-{2 \lambda z }} \,\,\,\, .
\end{eqnarray}
Considering eqs.\eqref{moddil}-\eqref{v1-1}  one sees that the mass term is the main responsible of  the  difference between the scalar and vector cases when the geometry is perturbed,
while its effect turns out to be the same when the background dilaton is modified.
The obtained potentials are depicted in Fig.\ref{fig:scalfig}.

Eq.\eqref{geneqpot} with the
new potentials \eqref{pert-pot} and \eqref{pert-pot-1}
 can be solved perturbatively, and  for small values of the parameter  $\l$
the spectra are modified:
\begin{equation}\label{pert-mass}
  m_n^2=m_{n,(0)}^2+\lambda m_{n,(1)}^2 \,\,\, .
\end{equation}

For the scalar glueball, modifying the dilaton field according to \eqref{moddil}
we obtain for the first three states:
\begin{eqnarray}
  m_{0}^2 & = & 8+\l\fr{3\sqrt{\pi}}{2}\non\\
  m_{1}^2 & = & 12+\l\fr{27\sqrt{\pi}}{16}\\
  m_{2}^2 & = & 16+\l\fr{237\sqrt{\pi}}{128} \,\,\, .\non
\end{eqnarray}

\noi On the other hand, modifying the geometry  according to \eqref{modgeo}  the masses of the  first three states  the  spectrum  are given by:
\begin{eqnarray}
  m_{0}^2 & = & 8+\l\fr{9\sqrt{\pi}}{2}\non\\
  m_{1}^2 & = & 12+\l\fr{81\sqrt{\pi}}{16}\\
  m_{2}^2 & = & 16+\l\fr{711\sqrt{\pi}}{128} \,\,\,\ .\non
\end{eqnarray}

\begin{figure}[ht]
 \begin{center}
\includegraphics[width=0.45\textwidth] {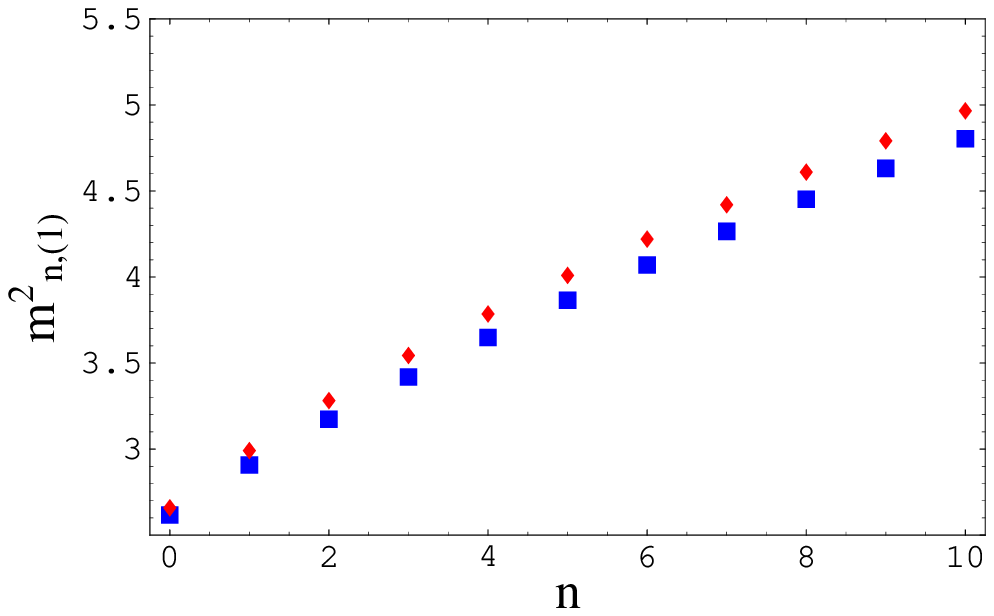}
\includegraphics[width=0.45\textwidth] {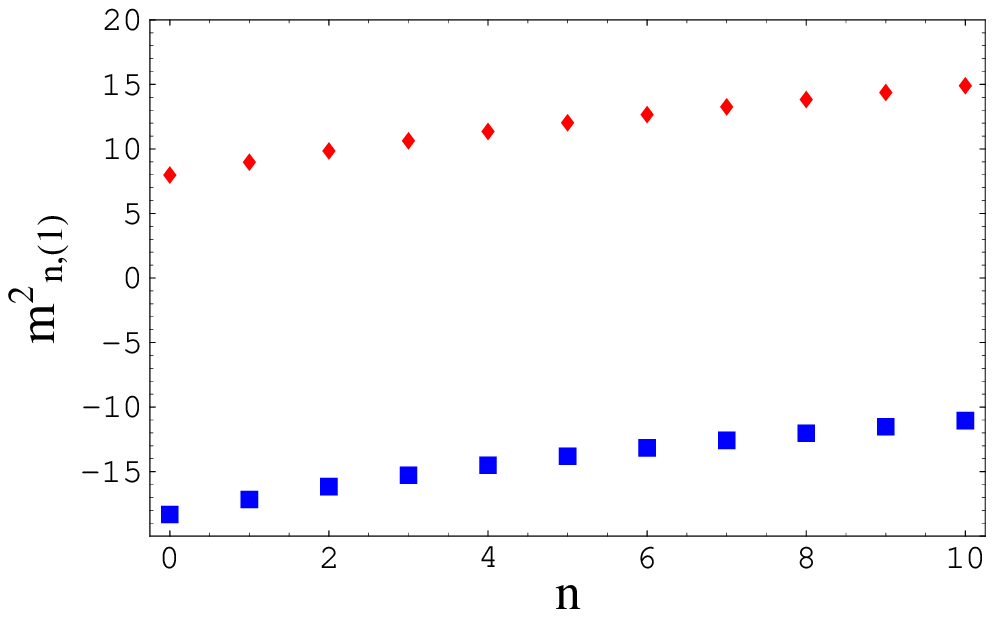}
 \end{center}
 \caption{\footnotesize{  Mass shifts
 in eq.\eqref{pert-mass} for scalar (red diamonds) and vector  glueballs (blue boxes) obtained by modifying the dilaton field  (left) or the metric function  (right).}}\label{fig:listfig}
\end{figure}

Also for vector glueballs a different  spectrum is
obtained, depending on the perturbations  \eqref{moddil} or \eqref{modgeo}.
Modifying the dilaton field the    values of the first
three states of the spectrum are:

\begin{eqnarray}
  m_{0}^2 & = & 12+\l\fr{189\sqrt{\pi}}{128}\non\\
  m_{1}^2 & = & 16+\l\fr{105\sqrt{\pi}}{64}\\
  m_{2}^2 & = & 20+\l\fr{14667\sqrt{\pi}}{8192}\non \,\,\, ,
\end{eqnarray}
while modifying  the geometry we obtain:

\begin{eqnarray}
  m_{0}^2 & = & 12-\l\fr{1323\sqrt{\pi}}{128}\non\\
  m_{1}^2 & = & 16-\l\fr{1239\sqrt{\pi}}{128}\\
  m_{2}^2 & = & 20-\l\fr{74685\sqrt{\pi}}{8192} \,\,.\non
\end{eqnarray}
Therefore,  when  the dilaton field is  modified,  the  mass shifts
have the same sign in  case of scalar and vector glueballs,
while the sign is opposite when the geometry is changed. This result
 is depicted  in Fig.\ref{fig:listfig}, where the mass shifts
$m_{n,(1)}^2$ defined in \eqref{pert-mass} are plotted  for the
first 11 states in case of  modified dilaton or  geometry.

Different predictions at ${\cal O}(\lambda)$ for the vector and scalar glueball  mass difference
are obtained   modifying  either the dilaton or the geometry.
 Modifying the dilaton, we get
\begin{equation}\label{res0}
m_{G_1}^2-m_{G_0}^2=4 -\frac{3 {\sqrt \pi}}{128} \l
\end{equation}
while, modifying the metric function,  we obtain:
\begin{equation}\label{res}
m_{G_1}^2-m_{G_0}^2=4-\frac{1899  {\sqrt \pi}}{128} \l  \,\,\,\, .
\end{equation}
Therefore, the mass splitting between vector and scalar glueballs increases if $\l$ is negative, and the
maximum effect is produced for the same value of  $\l$  when the metric function is perturbed.
 This can be considered as an indication on the type of constraints the background fields in the bulk must satisfy.

\section{Conclusions}
We have discussed how  the QCD holographic model proposed in  \cite{son2}, with the hard IR wall replaced by a 
background dilaton field, allows to predict  the  light glueball spectrum.  Scalar and pseudoscalar glueballs turn out to be degenerate if the fields representing the holographic correspondent of the
respective QCD operators are both massless zero forms.  Vector glueballs turn out to be heavier than the scalar ones, and
 the dependence of their masses on the radial quantum number is the same as obtained for  $\r$ and higher spin mesons. Combining the calculations of the glueball and  $\r$ masses in the same holographic model,  the glueballs  turn out to be lighter than predicted in other approaches.
 
 We have investigated how the masses change as a consequence of perturbing  the dilaton in the UV or the bulk geometry in the IR,  finding that 
 constraints in the  the bottom-up approach can be found if information on the spectra 
 from other approaches is considered.  Such constraints should be taken into account in the  attempt to   construct  the QCD gravitational dual. 

\vspace{1cm}

\noi {\bf Acknowledgments.}\\
\noi
We are grateful to D.T. ~Son and  L. ~Yaffe for discussions. We thank    F.~Canfora for information.  This work was supported in part by the EU Contract   No. MRTN-CT-2006-035482, "FLAVIAnet".

\end{document}